\documentstyle[prb,aps,psfig,multicol]{revtex}

\begin{document} 
 
\draft 

\title{Epitaxy and Magnetotransport of Sr$_{2}$FeMoO$_6$ thin films}

\author{W.~Westerburg\cite{byline}, D.~Reisinger, and G.~Jakob}  

\address{Institute of Physics, University of Mainz, 55099 Mainz, Germany} 

\date{27 Januar 2000} 

\maketitle 

\begin{abstract} 
By pulsed-laser deposition epitaxial thin films of Sr$_{2}$FeMoO$_6$ have been prepared on (100) SrTiO$_3$
substrates. Already for a deposition temperature of 320$^\circ$C epitaxial growth is achieved. 
Depending on deposition parameters the films show metallic or semiconducting behavior.
At high (low) deposition temperature the Fe,Mo sublattice has a rock-salt (random) structure.
The metallic samples have a large negative magnetoresistance which peaks at the Curie temperature. 
The magnetic moment was determined to 4\ $\mu_B$ per formula unit (f.u.), in agreement with the expected value for an ideal
ferrimagnetic arrangement. We found at 300\ K an ordinary Hall coefficient of $-6.01\times10^{-10}$\ m$^3$/As,
corresponding to an electronlike charge-carrier density of 1.3\ per Fe,Mo-pair.
In the semiconducting films the magnetic moment is reduced to 1\ $\mu_B$/f.u. due to disorder in the Fe,Mo sublattice. 
In low fields an anomalous holelike contribution dominates
the Hall voltage, which vanishes at low temperatures for the metallic films only. 
\end{abstract}  

\pacs{PACS numbers: 75.30.Vn, 73.50.Jt, 75.70.-i}  

%
\begin{multicols}{2}
The observation of colossal magnetoresistance in the half-metallic perovskite manganites has led to 
an intense research on ferromagnetic oxides. \cite{Coey99} 
Recently, a large room temperature (RT) magnetoresistance was found in Sr$_{2}$FeMoO$_6$ (SFMO)
\cite{Kobayashi98}, a material belonging to the class of double-perovskites ($AA'BB'$O$_6$). \cite{Anderson93}
Depending on the metal ion radius the $B$ and $B'$ ions arrange in a random or ordered fashion.
For the latter a layered or rock-salt structure is observed. 
The high Curie temperature and the high spin-polarization render these materials 
attractive as part of magnetic field sensors, e.g. in magnetic tunnel junctions. \cite{wester99}
The results reported lately on epitaxial thin film preparation on SrTiO$_3$ (STO) substrates 
are not consistent. Metallic as well as semiconducting behavior was found. \cite{Manako99,Asano99,Yin99} 
Also the magnetic saturation moments were smaller than expected suggesting disorder in the rock-salt arrangement. \cite{Ogale99}
Our goal was to find the preparation parameters for 
metallic, fully ordered films.  
We investigate in this letter in detail the 
differences between epitaxial metallic and semiconducting films with respect to their structural, magnetic and magnetotransport 
behavior, including Hall effect. We prepared a whole series of samples, but will discuss exhaustively 
two samples A and B which mark all the general differences.

%
SFMO thin films were prepared by pulsed laser ablation in an oxygen partial pressure of $10^{-1}-10^{-7}$\ Torr 
or in argon atmosphere of $10^{-1}$\ Torr from a stoichiometric target on (100) STO substrates.
During deposition the substrate temperature $T_{\mathrm D}$ was constant with values covering the range 
from 300$^\circ$C to 950$^\circ$C. Crystal structure investigations were performed using a 
two-circle and a four-circle X-ray diffractometer. 
The magnetic properties were determined with a SQUID magnetometer. 
With a Mireau interferometer we evaluated the film thicknesses to typically 100\ nm.
By standard photolithographic methods the samples were patterned to a 3\ mm wide and 8\ mm long bridge.
The longitudinal resistivity was measured by the standard four-point technique with a DC current. 
Below RT a standard superconducting
magnet and above RT a cryostat with a furnace in a RT bore were used. The procedure for Hall effect
measurements of the patterned samples is described in detail elsewhere. \cite{Jakob98}

%
The chemical composition of the films presented here was
identical to the nominal composition of the target (Sr$_{2}$FeMoO$_{6\pm\delta}$)
as determined by Rutherford backscattering on a reference sample on MgO.
Although the dimension of the crystallographic unit cell is $\sqrt{2}a_0\times\sqrt{2}a_0\times2a_0$,
where $a_0$ is the lattice parameter for a single $AB$O$_3$ perovskite ($a_0\approx$ 4\ \AA),
we use the larger cell doubled in all directions to underline the symmetry of the $B,B'$ rock-salt arrangement.
The films have a (00$l$) orientation perpendicular to the STO plane. 
A segregation into clusters of
compositions SrMoO$_3$ ($2a_0=7.950$\ \AA{})\cite{Brixner60} and SrFeO$_3$ ($2a_0=7.738$\ \AA{})\cite{Yakel55} 
should be visible in X-ray diffraction either as severe peak broadening for small clusters 
or as peak splitting for large clusters. Both effects are not observed. The
in-plane orientation, film axes parallel to substrate axes, was checked by $\phi$-scans of the symmetry
equivalent \{224\} reflections.

Both deposition temperature $T_{\mathrm D}$ and oxygen partial pressure play a crucial role for 
phase formation and epitaxy. Growth of SFMO films at oxygen partial pressures above $10^{-1}$\ Torr
was not possible irrespective of the substrate temperature, 
but a polycrystalline, yellow, insulating phase formed. Lower oxygen partial pressure 
during deposition stabilized formation of the SFMO phase. 
In the following we report the influence of the substrate temperature on phase formation and epitaxy 
for a series of films deposited at a very low oxygen partial pressure. This was realized either in 
flow of pure oxygen at a pressure of $10^{-5}$\ Torr or in flow of pure argon (99.996\%) at a total pressure
of  $10^{-1}$\ Torr. 
At $T_{\mathrm D}=320^\circ$C there is a sharp phase boundary
for the epitaxial thin film growth. 
For lower temperatures the same insulating yellow phase was observed. 
At 320$^\circ$C up to the highest temperatures achievable with our heater
of 950$^\circ$C the samples are single phase, dark and a high degree of $a,b$ and $c$-axis orientation is achieved.
Annealing these samples in an oxygen partial pressure of more than $10^{-1}$\ Torr leads again to the polycrystalline 
yellow phase.  
Already at 320$^\circ$C perfect epitaxial growth is obtained with rocking curve width $\Delta\omega$
of the (004) reflections 
below 0.04$^\circ$.
Compared to results gained on other related perovskites as manganites or high $T_c$ superconductors,
this is an astonishingly sharp crossover and low epitaxy temperature. 
A lowering of the epitaxy temperature will be expected for a perfect in-plane lattice match of film and substrate. 
However, this is not the case here.
The bulk material is cubic with a doubled perovskite unit cell. It was refined in the space group Fm3m with a 
lattice constant of 7.897\ \AA{}. \cite{garcia99}
For the sample A, prepared at $T_{\mathrm D}=320^\circ$C all axes, $a,b$ and $c$, 
are elongated to 7.972\ \AA{} and 8.057\ \AA, respectively. The lattice mismatch
to the substrate is 2\%. 
 
\begin{figure}[htp]
\psfig{file=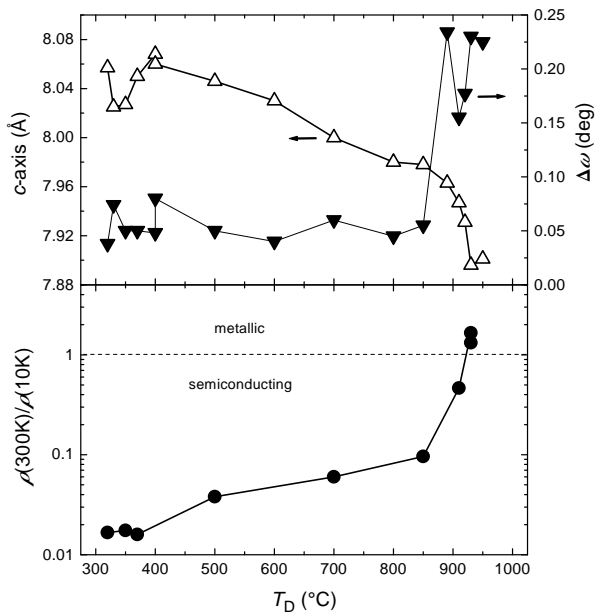,width=0.93\columnwidth} 
\vspace{0.5em} 
\caption{Upper panel: Dependence of $c$-axis length (left axis) and rocking curve widths $\Delta\omega$ of 
the (004) reflection (right axis) on deposition temperature $T_{\mathrm D}$; 
lower panel: Dependence of resistance ratio $\rho$(300\ K)/$\rho$(10\ K)
on substrate temperature dependence during fabrication.}
\label{CaxisvT}
\end{figure}

With a higher $T_{\mathrm D}$ the $c$-axis length decreases continuously while
the rocking curve widths $\Delta\omega$ of the (004) reflection remain  
around 0.05$^\circ$ for substrate temperatures 
$320\le T_{\mathrm D}\le 910^\circ$C, as can be seen in Fig.\ \ref{CaxisvT}.
For higher deposition temperatures the rocking curves 
broaden to an angular spread of 0.23$^\circ$ at 930$^\circ$C (sample B).
At this deposition temperature we obtained an in-plane lattice constant 
of 7.876\ \AA{} and an out-of-plane constant of 7.896\ \AA{}. Therefore the cell volume of 
sample B is decreased significantly by 4.5\% compared to sample A. 
Investigations of the reciprocal lattice show clear differences in the 
\{111\} reflections of the two films. They are present in sample B but absent in sample A. 
Intensity in this reflection is generated by the Fe,Mo ordered rock-salt arrangement.
The intensity ratio of the (111) and (004) reflection in sample B is close to 
the expected value in the tetragonal I4/mmm structure. 
In contrast the absence of this reflection in sample A indicates 
random $B,B'$ site occupation resulting in an enlarged cell volume.
The detailed dependence of the (111) intensity on $T_{\mathrm D}$ will be the subject of a future study.
In the following we show that the differences in the crystal structure go along with
respective differences in the transport properties.

\begin{figure}[htp]
\psfig{file=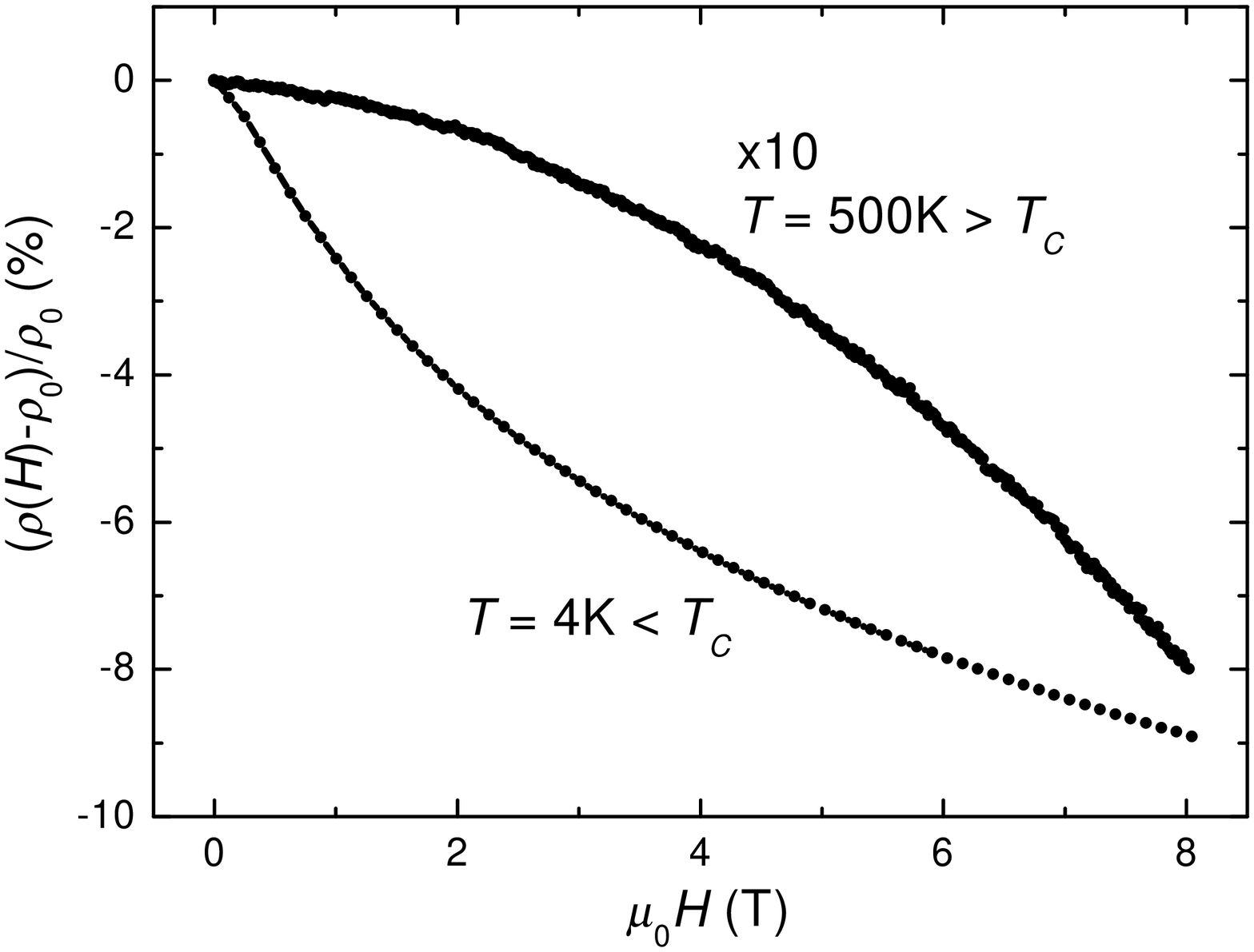,width=0.8\columnwidth} 
\vspace{0.5em} 
\caption{High field magnetoresistance below and above (multiplied by a factor 10) the Curie temperature.}
\label{RvB} 
\end{figure}

\begin{figure}[htp]
\psfig{file=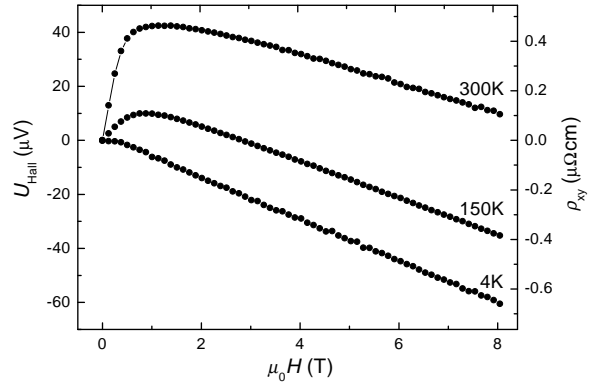,width=0.9\columnwidth} 
\vspace{0.5em} 
\caption{Measured Hall voltages $U_{{\mathrm Hall}}$ (left axis) and 
transverse resistivities $\rho_{xy}$ (right axis) as
functions of magnetic field for constant temperatures.}
\label{Hall} 
\end{figure}

\begin{figure}[htp]
\psfig{file=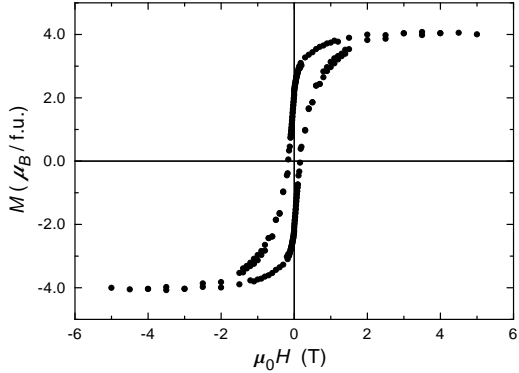,width=0.8\columnwidth} 
\vspace{0.5em} 
\caption{Ferromagnetic hysteresis curve for the SFMO film from -5\ T to 5\ T taken at 10\ K. A linear diamagnetic background 
from the substrate was subtracted from the raw data. 
The direction of $H$ was in the plane of the film.
}
\label{Hyst} 
\end{figure}

%
The temperature coefficient of the resistivity depends on deposition conditions. 
For deposition temperatures  $320\le T_{\mathrm D}\le 920^\circ$C the temperature coefficient is negative, 
i.e. semiconductorlike behavior.
\begin{figure}[htp]
\psfig{file=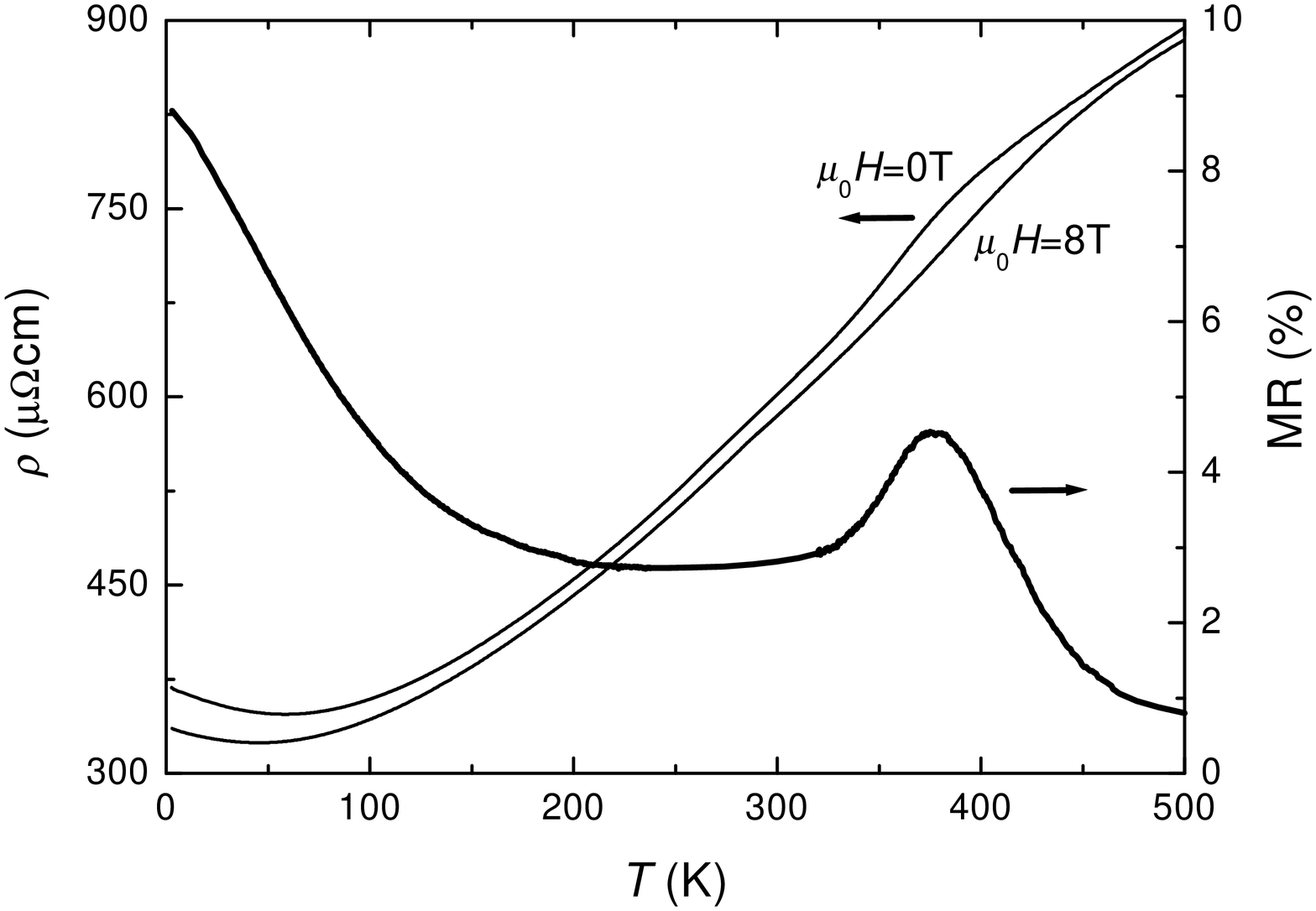,width=0.9\columnwidth} 
\vspace{0.5em} 
\caption{Temperature dependent resistivity in zero field and in 8\ T magnetic field
(left axis) and temperature dependence of negative magnetoresistance 
MR$=[\rho(T,\mu_0 H=0{\mathrm\ T})-\rho(T,\mu_0 H=8{\mathrm\ T})]/\rho(T,\mu_0 H=0{\mathrm\ T})$ 
(right axis).}
\label{RvT} 
\end{figure}
 The transport properties for these semiconducting films are described in detail elsewhere. \cite{wester00} 
Here we give in short the values for one of these films, sample A, whose crystallographic properties are 
discussed above. Its resistivity increases almost two orders of magnitude while cooling from RT down to 10\ K. 
No significant MR is visible in this sample which marks the lowest 
possible deposition temperature for epitaxial film growth. 
With increasing deposition temperatures the ratios between the RT resistivities and the 
respective low temperature resistivities ($T=10$\ K) increase, as is shown in 
Fig.\ \ref{CaxisvT}. At 930$^\circ$C finally the ratio is larger than unity, i.e.\ a metallic behavior occurs. 
The transition from a negative to positive temperature coefficient of the resistivity is connected 
with the appearance of the (111) reflection signaling $B,B'$-site ordering. 
Figure \ref{RvT} presents the resistivity (left axis) of the metallic sample as a function of temperature 
in zero field and in high magnetic field of 8\ T. A positive temperature coefficient exists in the temperature range 
from 50\ K to 500\ K. Below 50\ K a small increase can be observed with a residual resistivity at 
4\ K of 360\ $\mu\Omega${}cm.  
The magnetic ordering shows up in the temperature dependence of the resistivity as a small anomaly in the zero 
field curve near the Curie temperature $T_C$. The anomaly vanishes by applying a
high magnetic field. This temperature dependence is similar to the Sr-doped manganites which have the highest
$T_C$ within the manganites. \cite{West99}\\
Two different regimes with a high magnetoresistance MR, defined by 
MR=$[\rho(T,\mu_0H=0{\mathrm T})-\rho(T,\mu_0H=8{\mathrm T})]/\rho(T,\mu_0H=0{\mathrm T})$ (right axis) can be
distinguished. At very low temperatures the MR increases supposedly due to grain boundary effects. \cite{Maignan99} 
The peak at 380\ K in the neighbourhood of the Curie temperature $T_C$ is probably due
to the suppression of spin-fluctuations by the external magnetic field.
These results are similar to those of polycrystalline manganites. \cite{Martin99}
The magnetic field dependence of the resistivity below and above (multiplied by a factor 10) $T_C$ is shown in 
Fig.\ \ref{RvB}. 
Both curves have opposite curvatures.
The MR is for small fields proportional to $H^2$ for $T>T_C$ while for $T<T_C$ a linear dependence is found 
in agreement with the symmetry considerations for the manganites by Snyder {\it et al.} \cite{Snyder69}\\
The Hall effect in a ferromagnetic material is described by
\begin{equation}
	\label{RhoHall}
	\rho_{xy}=R_H B+R_A\mu_0 M
\label{Hallgl}
\end{equation}
with the the magnetization $M$ and the ordinary and anomalous Hall coefficients $R_H$ and 
$R_A$, respectively. \cite{Karplus54}
We evaluate in this paper the high field regime where the magnetization is close to saturation 
and ${\mathrm d}\rho_{xy}/{\mathrm d}B\approx{\mathrm d}\rho_{xy}/{\mathrm d}(\mu_0 H)$.
We measured the Hall voltage $U_{{\mathrm Hall}}$ with a current of 1\ mA at constant temperatures $T=4,150,300$\ K in 
magnetic fields up to 8\ T. 
The results for sample B are shown in Fig.\ \ref{Hall}.
 In this figure the symbol sizes are larger than the experimental errors. 
With increasing temperature a steep increase
of $U_{{\mathrm Hall}}$ in low fields can be seen. At 1\ T a maximum exists and at higher fields a linear 
negative slope is visible. This behavior is typical for ferromagnets where the anomalous Hall effect
dominates due to the change in magnetization in low fields. At higher fields the magnetization saturates. 
Then according to Eq.\ \ref{Hallgl} 
the anomalous Hall contribution is constant and the ordinary Hall effect is apparent. 
In contrast to the manganites $R_A$ is holelike and $R_H$ electronlike. \cite{Jakob98}
The same sign correlation as in SFMO was also observed in iron and ferromagnetic iron alloys. \cite{Bergmann79}
In the following we focus on the discussion of the linear slopes ${\mathrm d}\rho_{xy}/{\mathrm d}(\mu_0 H)$ 
in the high-field regime, given only by the ordinary Hall contribution.
We find at RT a Hall coefficent of $-6.01\times10^{-10}$\ m$^3$/As, corresponding to $n_e=1.3$ electrons per $B,B'$-pair
in a single-band model. 
With decreasing temperature $n_e$ decreases slightly to 0.9 electrons per f.u.
The zero field mobility at 4\ K and RT is calculated with these values of the charge-carrier density to 
232\ mm$^2$/Vs and 100\ mm$^2$/Vs, respectively. 
A second metallic sample has almost the same charge-carrier concentration.
However, in the semiconducting films a higher $n_e$ of
4.1 electrons per $B,B'$-pair is measured and it increases with decreasing temperature. \cite{wester00}
The increase of the nominal charge-carrier density in the semiconducting sample compared to the metallic 
one and the high value shows, that a single band model is not appropriate for this compound and 
band structure effects dominate the Hall effect. Nevertheless, we give the nominal charge-carrier concentration
per f.u. for easier comparison with other reports.
The values gained from the metallic sample B are similar to single crystal data
by Tomioka {\it et al.} \cite{Tomioka00} 
At RT we find an anomalous Hall coefficient $R_A$=$1.56\times10^{-8}$\ m$^3$/As
which vanishes for our metallic SFMO films at very low temperatures. This is 
expected by theory for an undistorted ferromagnetic state. \cite{Smit55}
SQUID magnetization measurements confirm this picture showing full saturation near 4.0\ $\mu_B$/f.u.
The nonvanishing anomalous Hall contribution in our semiconducting SFMO thin films \cite{wester00} and the  
single crystal data \cite{Tomioka00} is an indication for a distorted ferrimagnetic state, 
which is indicated also by the reduced magnetizations of these samples.
The values of the saturation magnetization $M_S^{\mathrm exp}$ are  1.0\ $\mu_B$/f.u. for the semiconducting SFMO film 
and  3.2\ $\mu_B$/f.u. for the metallic single crystal.
The reduction of the magnetic moment is due to $B,B'$-site disorder 
causing antiferromagnetic Fe-O-Fe bonds. \cite{Ogale99} 
M\"o\ss{}bauer measurements suggest a Fe$^{3+}$(S=5/2) configuration and with Mo$^{5+}$(S=1/2) one
expects in a ferrimagnetic arrangement a value of 4\ $\mu_B$/f.u. \cite{Pinsard00} 
In Fig.\ \ref{Hyst} we show the hysteresis loop
at 10\ K for the metallic sample B. 
The saturation magnetization $M_S^{\mathrm exp}$ is close to the ideal value 
of $M_S^{\mathrm theo}=4$\ $\mu_B$/f.u.

In Table \ref{proper} crystal structure, magnetotransport, and magnetization data 
of the samples A and B are listed to emphasize the
differences between low temperature deposited SFMO thin films
and their high temperature deposited counterparts.

%
Concluding we prepared epitaxial Sr$_{2}$FeMoO$_6$ thin films with a high degree of epitaxy by pulsed laser
deposition. Already for a substrate temperature during fabrication of 320$^\circ$C epitaxial growth is achieved. 
Depending on deposition parameters we found a strong variation in cell volume and degree of order of the 
$B,B'$-site occupation, resulting in metallic or semiconducting behavior of the resistivity.
For the metallic samples the negative magnetoresistance peaks in the neighbourhood of 
the Curie temperature. The magnetic moment is equal 
to the expected value for a ferrimagnetic arrangement of the Fe and Mo cations. At low temperatures
the anomalous Hall effect
vanishes, indicating an undistorted ferrimagnetic state in agreement with magnetization
measurements. 
The ordinary and anomalous Hall coefficients are negative and positive, respectively.
Both coefficients have reversed sign 
compared to the colossal magnetoresistive manganites. We found a charge-carrier
concentration of 1.3 electrons per f.u. at room temperature.

%
The authors thank P. Latorre Carmona for experimental assistance, P.~G\"utlich from Institut f\"ur Anorganische Chemie 
und Analytische Chemie for usage of the magnet cryostat with RT access and G. Linker from 
Forschungszentrum Karlsruhe for the Rutherford backscattering analysis
of the film stoichiometry.  
This work was supported by the Deutsche Forschungsgemeinschaft through Project No. JA821/1-3 and the
Materialwissenschaftlichen Forschungszentrum (MWFZ) Mainz.

\begin{table}[htp]
\caption{Differences in physical properties of sample A (semiconducting) and sample B (metallic).}
\begin{tabular}{|c|c|c|}
Sample & A & B \\ \hline
$T_{\mathrm D}$ ($^\circ$C) & 320 & 930 \\
$a,b$-axes (\AA{}) & 7.972 & 7.876 \\
$c$-axis (\AA{}) & 8.057 & 7.896 \\
cell volume (\AA{}$^3$) & 512.045 & 489.800 \\
(111) reflection & absent & present \\
$\rho$(300\ K)/$\rho$(10\ K) & $<< 1$ & $> 1$\\
negative MR & absent & present \\
$R_H$(300\ K) (10$^{-10}$\ m$^3$/As)& -1.87 & -6.01\\
$R_A$(4\ K) & present & absent\\
$M_S^{\mathrm exp}$/$M_S^{\mathrm theo}$ & 0.25 & 1\\ 
\end{tabular}
\label{proper}
\end{table}

\end{multicols}
\end{document}